\documentstyle[12pt]{article}

         \def\la{\lambda}
         \def\be{\begin{equation}}
         \def\bea{\begin{eqnarray}}
         
         \def\o{\over}
         \def\ep{\epsilon}

         \def\ee{\end{equation}}
         \def\eea{\end{eqnarray}}
         \def\R{\rm {I\kern-.200em R}}
         \def\C{\rm {I\kern-.520em C}}
         \def\c{\chi}

\begin{titlepage}
\begin{document}
\vspace*{5mm}
\begin{center} {\Large \bf $n$-point functions of $2d$ Yang-Mills\\
\vskip 0.35cm
theories on Riemann surfaces}\\
\vskip 1cm
M.Alimohammadi$ ^{a,b}$ \footnote {e-mail:alimohmd@rose.ipm.ac.ir} and
M.Khorrami$ ^{a,b,c}$ \footnote {e-mail:mamwad@rose.ipm.ac.ir}\\
\vskip 1cm
{\it $^a$ Institute for Studies in Theoretical Physics and Mathematics, }\\
{\it P.O.Box 19395-5746, Tehran, Iran}\\
{\it $^b$ Physics Department, University of Teheran, North Karegar,} \\
{\it Tehran, Iran }\\
{\it $^c$ Institute for Advanced Studies in Basic Physics , P.O.Box 159 ,}\\
{\it  Gava Zang , Zanjan 45195 , Iran }\\
\end{center}
\vskip 2cm
\begin{abstract}

Using the simple path integral method we calculate the $n$-point functions of
field strength of Yang-Mills theories on arbitrary two-dimensional Riemann
surfaces. In $U(1)$ case we show that the correlators consist of two parts ,
a free and an $x$-independent part. In the case of non-abelian semisimple compact
gauge groups we find the non-gauge invariant correlators in Schwinger-Fock gauge
and show that it is also divided to a free and an almost $x$-independent part.
We also find the gauge-invariant Green functions and show that they
correspond to a free field theory.
\end{abstract}
\newpage
{\bf Introduction}

\hskip 0.25cm

In recent years there have been many efforts to understand the two dimensional
Yang-Mills theories. The partition function of these theories on $\Sigma_g$ ,
a two-dimensional Riemann surface of genus $g$ , has been calculated in the
context of lattice gauge theory $[1,2]$ . On the other hand the string
interpretation of $2d$ Yang-Mills theory was discussed in $[3]$ and $[4]$
by studying the $1/N$ expansion of the partition function for $SU(N)$ gauge
group. It was shown that the coefficients of this expansion are determined by
a sum over maps from a two-dimensional surface onto the two-dimensional target
space.

Two-dimensinal Yang-Mills theories have also been studied by means of path
integral method $[5,6]$. In $[7]$ and $[8]$ the partition function and the
expectation values of Wilson loops have been calculated and in $[9-11]$ these
quantities were calculated by using the abelianization technique.

In this paper we study the correlation functions of field strengths of $2d$
Yang-Mills theories by path integral method in a simple way. We drive all $n$-point
functions on arbitrary surface. In the first part we consider the $U(1)$ gauge
group and by calculating $Z[J]$ , the partition function in the presence of an
external source , we compute the gauge-invariant correlators on $\Sigma_g$. We
show that the results consist of two parts , a part which comes from
a free field theory and a second part which is independent
of coordinates of fields. We also redrive the results by means of the expectation
value of Wilson loops.

In the second part , we investigate the non-abelian gauge theories. Using the
fermionic path integral representation of the trace of Wilson loops , we find the
$Z[J]$ in Schwinger-Fock gauge and calculate the angle ordered $n-$point
functions. We see that the non-gauge invariant correlators consist of a gauge-invariant
part , and another part which is almost independent of the coordinates of the fields.
We also extract the gauge-invariant part of the correlators and show that they
correspond to a free theory. At the end we justify our results by using the
Wilson loop correlators.

When this paper was nearly finished , we became aware of the preprint $[13]$
in which the two and four-point functions of $U(N)$ gauge group have been derived by
lengthy abelianization metod.

\hskip 1cm

{\bf 1 - Maxwell theory}

\hskip 0.25cm

Consider the partition function $Z[J]$ on $\Sigma_g$ :
\be Z[J]=\int {\cal D} \xi e^{-{1 \o 2\epsilon}\int \xi^2d\mu+\int\xi J d\mu}
\delta^p({1\over 2\pi}\int \xi d\mu) \ee
where the scalar field $\xi (x)$ is defined by $F_{\mu \nu}=\xi (x) \epsilon_{\mu \nu} ,
d\mu=\sqrt {g(x)}d^2x$ and
\be \delta^p({1\over 2\pi}\int \xi d\mu) =\sum_n\delta ({1\over 2\pi}\int \xi d\mu
-n)=\sum_ne^{in \int \xi d\mu}. \ee
The insertion of $\delta^p({1\over 2\pi}\int \xi d\mu)$ ensures that we are not
integrating over arbitrary two-forms but over curvatures of connections. Performing
the Gaussian integral $(1)$ , we find
\be Z[J]=Z_1[J]Z_2[J] , \ee
where :
\be Z_1[J]=e^{{\epsilon \o 2}\int J^2 d\mu}\ee
which is the partition function of a free field theory , and
\be Z_2[J]=\sum_n {\rm exp}[in\epsilon \int J d\mu -{\epsilon \o 2}n^2A(\Sigma_g)]\ee
in which $A(\Sigma_g)$ is the area of $\Sigma_g$. The gauge-invariant correlators
of $\xi (x)$ is defined via :
\be<\xi (x_1) ... \xi (x_n)>={1 \o Z[0] }{\delta \o \delta J(x_1) } ...
{\delta \o \delta J(x_n) }Z[J]\vert_{J=0}\ee
Now $Z_1$ is the partition function of a free theory , so that its $n$-point
functions are zero unless $n$ is even. The $2n$-point functions factorize to the
two point function
\be G(x,y)=\ep \delta (x-y) .\ee
So
\be <\xi (x_1) ... \xi (x_{2n})>_1=\sum_p G(x_{i_1},x_{i_2})...
G(x_{i_{2n-1}},x_{i_{2n}}),\ee
where the summation is over all distinct pairings of the $2n$ indeces with
${(2n)! \o {(2!)^nn!}}$ terms.

The correlators corresponding to $Z_2$ are also calculated to be
\be <\xi (x_1) ... \xi (x_{2n})>_2={1 \o Z[0]}\sum_m(im\ep)^{2n}e^{-{\ep \o 2}
m^2A(\Sigma_g)},\ee
and the odd-point functions are zero. The complete $n$-point function is simply
obtained , using these two correlators. The odd-point functions are zero and the
even-point functions are :
\be<\xi (x_1) ... \xi (x_{2n})>=\sum_{m=0}^{2n}\sum_c<\xi (x_1) ... \xi (x_m)>_1
<\xi (x_{m+1}) ... \xi (x_{2n})>_2,\ee
where the inner summation is over all different ways of choosing $m$ indeces
from $2n$ indeces. As mentioned in the introduction the correlators consist
of a free and an $x$-independent parts.

Now it is useful to reproduce the above results from another method , that is
using the expectation value $<e^{i\alpha \oint_\gamma A}>$ where $\gamma$ is a
homologically trivial loop on $\Sigma_g$ :
$$<e^{i\alpha \oint_\gamma A}>={1\o Z[0]}
\int {\cal D} \xi e^{-{1 \o 2\epsilon}\int \xi^2d\mu+i\alpha\int_D\xi d\mu}
\delta^p({1\over 2\pi}\int \xi d\mu) $$
$$={1\o Z[0]}\sum_nexp\{ -{\ep \o 2}[n^2A(\Sigma_g)+\alpha^2A(D)+2\alpha n A(D)]\}$$
\be =1+\alpha^2{\ep /2 \o Z[0]}\sum_n(\ep n^2A^2(D)-A(D))e^{-{\ep \o 2}n^2
A(\Sigma_g)}+o(\alpha^2),\ee
where $A(D)$ is the area of disk $D$ , the boundary of which is $\gamma$. Expanding
the left hand side of eq.(11) in terms of $\alpha$ , gives
\be <e^{i\alpha \int_D A}>=1+i\alpha\int_D<\xi (x)>d\mu-{\alpha^2 \o 2}\int_D
<\xi (x) \xi (y) > d\mu (x) d\mu (y)+o(\alpha^2).\ee
By comparing the two side of eq.(11) , it is seen that $<\xi (x)>=0$ . In order
to find the two point function we use the following ansatz
\be <\xi (x) \xi (y) >=M\delta (x-y) +N .\ee
In fact , as the theory is topological , it is reasonable that the two point
function consists of an $x$-independent term and a term which just sees if the
two point are equal or not. This term should be a delta term , because of the
Gaussian nature of the integrand in the partition function. Using this ansatz ,
it is readily seen that
\be<\xi (x) \xi (y) >=\ep \delta (x-y) - {\ep^2 \o Z[0]}\sum_mm^2e^{-{\ep \o 2}
m^2A(\Sigma_g)},\ee
which is the result obtained previously. Other $n$-point functions can also
be obtained in this way.

\hskip 1cm

{\bf 2 - Yang-Mills theory}

\hskip 0.25cm

In this section we are going to calculate $Z[J]$ for a non-abelian semisimple
compact gauge group $G$. To begin , we consider the wavefunction $\psi_D[J]$
on the disk $D$. If $\gamma$ is the boundary of $D$ , $\gamma = \partial D$
, we choose the boundary condition to be ${\rm Pexp}{\oint_\gamma A}=g_1\in G$ (modulo
conjugation) , and therefore $\psi_D[J]$ is defined as :
\be \psi_D[J]=\int {\cal D} \xi e^{-{1 \o 2\epsilon}\int \xi^a\xi_ad\mu+\int\xi^a J_a d\mu}
\delta({\rm Pexp}\oint_\gamma A,g_1). \ee
We have
\be \delta (h,g_1)=\sum_{\la \in {\hat G}}\c_\la (h)\c_\la (g_1^{-1}) \ee
where the summation is over all irreducible unitary representation of the group ,
 and $\c_\la $ is the character of the representation. We then use the fermionic
path integral representation of the Wilson loop $[7,12]$
\be \c_\la (Pexp\oint_\gamma A)=\int {\cal D}\eta {\cal D}{\bar \eta}e^{
\int_0^1 dt {\bar \eta}(t){\dot \eta}(t)+\oint_\gamma {\bar \eta}(t)A(t)\eta (t)}
{\bar \eta}(1)\eta(0), \ee
where $\eta$ is a Grassmann valued vector in the representation $\la $.We also
use the Schwinger-Fock gauge:
\be A_\mu^a(x)=\int_0^1dssx^\nu F^a_{\nu \mu}(sx). \ee
In this gauge , one can write
\be \oint_\gamma {\bar \eta}(t)A(t)\eta (t)=\int_D{\bar \eta}F\eta .\ee
Using this , it is easily seen that
\be \psi_D[J]=\sum_\la \c_\la (g_1^{-1})\int {\cal D}\eta {\cal D}{\bar \eta}e^{
\int_0^1 dt {\bar \eta}(t){\dot \eta}(t)}e^{{\ep \o 2}\int
(J^a+{\bar \eta}T_\la^a\eta ) (J_a+{\bar \eta}T_{a\la}\eta ){\sqrt g}dsdt}
{\bar \eta}(1)\eta(0), \ee
where $T_{a,\la}$'s are the generators of the group in the representation $\la$.
This integral is also calculated to be ( see the appendix of $[8]$ )
\be \psi_D[J]=Z_1[J]\psi_{2,D}[J] ,\ee
where
\be Z_1[J]=e^{{\ep \o 2}\int J^aJ_ad\mu} ,\ee
and
\be \psi_{2,D}[J]=\sum_\la \c_\la (g_1^{-1})e^{-{\ep \o 2}c_2(\la)A(D)}
\c_\la ({\cal P}{\rm exp} \ep \int dt \int ds {\sqrt g}J(s,t)) .\ee
Here the ordering is according to $t$ ( the angle coordinate ). The disk is
parametrized by the coordinates $s$ ( the radial coordinate ) and $t$ ( the
angle coordinate ) , and $c_2(\la)$ is the quadratic Casimir of the representation
$\la $.

It is now easy to see that
\be < \xi^{a_1}(x_1)...\xi^{a_n}(x_n)>_{2,D}
= {1 \o Z_D[0]}\sum_\la \c_\la (g_1^{-1})e^{-{\ep \o 2}c_2(\la)A(D)}\ep^n\c_\la (
T^{a_1}...T^{a_n})\ee
$ {\rm for} \ \ \ t(x_1)<...<t(x_n) $.

To find the correlators on an arbitrary closed surface , it suffices to glue
the disk to $\Sigma_{g,1}$, a genus-$g$ surface with a boundary , with boundary
condition ${\rm Pexp} \oint_\gamma A=g_1^{-1} $ :
\be < \xi^{a_1}(x_1)...\xi^{a_n}(x_n)>_{2,\Sigma_g}={1 \o Z_{\Sigma_g}[0]}
\int dg_1< \xi^{a_1}(x_1)...\xi^{a_n}(x_n)>_{2,D}\psi_{\Sigma_{g,1}}(g_1^{-1}),\ee
where ( from $[7]$ ) we have
\be Z_{\Sigma_g}[0]=\sum_\la d(\la)^{2-2g}e^{-{\ep \o 2}c_2(\la)A(\Sigma_g)},\ee
in which $d(\la )$ is the dimension of the representation $\la$ , and
\be \psi_{\Sigma_{g,1}}(g_1^{-1})=\sum_\la d(\la)^{2-2g-1}\c_\la (g_1)
e^{-{\ep \o 2}c_2(\la)A(\Sigma_{g,1})}.\ee
Using the orthogonality relation
\be \int \c_\la (g) \c_\mu (g^{-1})dg=\delta_{\la \mu } , \ee
one finds
\be < \xi^{a_1}(x_1)...\xi^{a_n}(x_n)>_{2,\Sigma_g}={1 \o Z_{\Sigma_g}[0]}
\sum_\la d(\la)^{2-2g-1}e^{-{\ep \o 2}c_2(\la)A(\Sigma_g)}\c_\la (T^{a_1}...
T^{a_n}),\ee
where it is understood that $t_1<...<t_n$ . This result obviously depends on
the choice of the coordinates and hence is not gauge invaraint.

$Z_1$ is the partition function of a free theory. So again we have ( like eq.(8) )
\be < \xi^{a_1}(x_1)...\xi^{a_n}(x_{2n})>_{1,\Sigma_g}=\sum_p
G^{a_{i_{1}}a_{i_{2}}}(x_{i_1},x_{i_2})...
G^{a_{i_{2n-1}}a_{i_{2n}}}(x_{i_{2n-1}},x_{i_{2n}})\ee
where
\be G^{ab}(x,y)=\ep \delta^{ab}\delta (x-y) .\ee
This completes the expression of the correlators of the Yang-Mills theory on
$\Sigma_g$ . Again they consist of a free part and a part which is almost
$x$-independent , that is , it depends only on the angular ordering of the coordinates.

Now the important question that arises is that which part of the above results are
gauge-invariant. First notice that the wavefunction (23) is not gauge-invariant.
This can be checked by noting that it is not invariant under the transformation
$J(x) \rightarrow U(x)J(x)U^{-1}(x)$ ( which induces the gauge transformation )
, because of the character term $\c_\la ( {\cal P}{\rm exp}{\ep \int J})$ . Now try to divide the
disk $D$ to $N$ parts and consider the wavefunction (23) for the disk
$D_N=D/N$ , ( with area $A/N$ ). Then try to find the wavefuction of the disk $D$
by gluing the wavefunction of the small disks. This is justified only if the
wavefunction is gauge invariant , because the radial and angle variables in each
disk is not the same as the other disks. However , as $N$ tends to infinity , it
is enough to calculate the wavefunction of the small disks only up to first order
of ($A/N$) . But , up to first order , we have
\be \psi_{2,D/N}[J]=
\sum_\la \c_\la(g_1^{-1})e^{-{\ep \o 2}c_2(\la)A({D\o N})}d(\la ), \ee
which is gauge invariant. Gluing these we find
\be Z_{\Sigma_g}^{G.I.}[J]=
e^{{\ep \o 2}\int J^aJ_ad\mu}Z_{\Sigma_g}[0].\ee
Therefore the gauge invariant part of the correlators are those quoted in eq.(30)
which are free.

Another method of calculating the gauge-invariant correlators is using the
expectation value of Wilson loops ( which are gauge-invariant ) $[7]$ :
\be <\c_\mu ({\rm Pexp} \oint_\gamma A)>={1 \o Z_{\Sigma_g}[0]}\sum_\la \sum_{\rho \in
\la \otimes \mu}d(\la )d(\rho )^{1-2g} exp\{-\ep [c_2(\la )A(D)+c_2(\rho )(A(
\Sigma_g)-A(D))] \} \ee
where $\gamma =\partial D$ . If , by symmetry consideration , we use the following
ansatz for the gauge-invariant two point function :
\be <\xi^a(x)\xi^b(y)>^{G.I.}=M\delta^{ab}\delta (x-y) , \ee
and , for small $A(D)$ , equate the linear term of both sides of eq.(34) , we will
find :
\be M=-{\ep \o d(\mu )c_2(\mu ) Z_{\Sigma_g}[0]}\sum_\la \sum_{\rho \in
\la \otimes \mu}d(\la )d(\rho )^{1-2g} e^{-{\ep \o 2}c_2(\rho )A(
\Sigma_g)}(c_2(\rho )-c_2(\la )). \ee
Then , using the identities :
$$ \sum_{\rho \in \la \otimes \mu}d(\rho )c_2(\rho )=d(\la )d(\mu )(c_2(\la )+
c_2(\mu ))$$
\be \sum_{\rho \in \la \otimes \mu}d(\rho )=d(\la )d(\mu ),\ee
it is seen that $M=\ep $ , which is consistent with (31).

\hskip 1cm

{\bf Acknowledgement}

\hskip 0.25cm

We would like to thank the research vice-chancellor of the university of Tehran.
\pagebreak

\end{titlepage}
\end{document}